\documentclass[aps,preprint]{revtex4}
\usepackage{epsfig}
\usepackage{amsmath}
\usepackage{float}
\usepackage{epsfig}

\begin{document}
\title
{\bf Coherent perfect absorption with and without lasing in  complex potentials}    
\author{Zafar Ahmed}
\email{zahmed@barc.gov.in}
\affiliation{Nuclear Physics Division, Bhabha Atomic Research Centre,Mumbai 400 085, India}
\date{\today}
\begin{abstract}
We prove that coherent perfect absorption (CPA) without lasing  is {\bf not} possible in the PT-symmetric domain as the s-matrix is such that $|\det S(\pm k)|=1$. We study coherent scattering from three complex potentials, one solved analytically and the other two numerically. We conjecture that in the domain of unbroken symmetry (when the potential has real discrete spectrum) neither spectral singularity nor CPA can occur. We show that Scarf II potential is a special model that can analytically and explicitly exhibit these as well as other novel phenomena and their subtleties. 
\end{abstract}
\maketitle
Recently, there has been a considerable revival of interest [1-19] in the scattering from complex one dimensional potentials, or equivalently, in the wave propagation through mediums with complex refractive index. Phenomena like (handedness) non-reciprocity [1] of reflectivity when the potential is spatially asymmetric, spectral singularity (SS) [real energy poles in reflection and transmission coefficients] [7]. The complex Scarf II [2,23] potential by virtue of its analytic amenability has helped in bringing [8,14] out spectral singularity and the related features explicitly.  

More interesting phenomena are revealed when two identical (coherent) beams are injected on the potential from both left and right. These phenomena are  coherent  perfect absorption (CPA)[9,12] and CPA with [10] lasing (spectral singularity).
Very intuitive ideas [7,9,10] have revealed these subtle phenomena and these occur as a possibility and not as a necessity in various domains of  complex scattering potentials. Thus, it  becomes an involved investigation of various domains and various parametric regimes in a complex potential in order to observe these effects. So far the analytically tractable and versatile Scarf II potential has not been utilized in this regard.  

Here in this paper, we prove that CPA without lasing  can not occur in PT-symmetric domains. We show that the exactly solvable complex Scarf II potential entails and exhibits these phenomena analytically and explicitly by bringing out their subtle features. Numerically solved examples are shown to follow the similar results as that of Scarf II. We also find that  in the domains where PT-symmetry is unbroken, spectral singularity and hence CPA does not occur. 

When a one-dimensional complex potential (vanishing asymptotically) is spatially asymmetric, the reflectivity is sensitive to the side of
incidence of wave whether it is left or right. It has been proved that [1]  
\begin{equation}
T_{left}(k)=T_{right}(k)=T(k) ~ \mbox{but}~
R_{left}(k) \ne R_{right}(k),
\end{equation}
Also see Refs. [3-7].
Following the same proof it has also been proposed [14] that for PT-symmetric potentials 
\begin{equation}
T(-k) = T(k), ~ \mbox{and}~ R_{left}(-k)=R_{right}(k).
\end{equation}
Recently these proposals have also been proved [19].
Here P means Parity transformation $(x \rightarrow -x)$ and T means Time reversal ($i \rightarrow -i, k \rightarrow -k$), $k$ is wave-number defined as $k=\sqrt{E}~(\hbar^2=1=2m)$.
For non-PT-symmetric cases we have [14]
\begin{equation}
T(-k) \ne T(k), R(-k) \ne R(k)~ \mbox{and}~ R_{left}(-k) \ne R_{right}(k).
\end{equation}
Let $r$  and $t$ be reflection and transmission (complex) amplitudes with phases as $\phi$ and $\theta$, respectively.
Then reflectivity, $R=|r|^2$, and transmittance, $T=|t|^2$. For complex PT-symmetric structures, it has been proved that [13]
\begin{equation}
\theta-\phi_{left}=\pi/2=\theta-\phi_{right}, ~\mbox{if}~T<1 ~ \mbox{and} \quad \theta-\phi_{left}=\pi/2=\phi_{right}-\theta,~\mbox{if}~ T>1.
\end{equation}
When two waves identical (coherent) in all respects are incident on a complex scattering potential from left and  right the 
the  $S$-matrix is given as [7]
\begin{equation}
S= \left (\begin{array} {cc}t & r_{left}  \\   r_{right} & t  \\ \end{array} \right),  \quad
\det S=t^2-r_{left}r_{right}
\end{equation}
Using Eq. (4) in (5) we find  that
\begin{equation}
|\det S|= T \pm \sqrt{R_{left} R_{right}}=1,
\end{equation}
following sub(super)-unitarity as proposed in [13,15].
In Hermitian quantum mechanics it is known that $( \phi_{left} - \theta )+(\phi_{right} -\theta)=\pi$ [16], for coherent injection at a  Hermitian potential the scattering matrix admits  $|S|=T+R=1$. Therefore, this  is yet another common feature shared by complex PT-symmetric potentials with Hermitian potentials.

On the contrary, the condition for CPA without lasing is [9]
\begin{equation}
|\det S(k_c)|=0.
\end{equation}
at a real positive energy, $E_c=k_c^2$. One can therefore see the impossibility of CPA for complex PT-symmetric cases, in view of
the result (6). 
 
However, the novel possibility of PT-symmetric potentials to display CPA with lasing  is distinct and different. It happens when at $E=E_*$, $T$ becomes infinity and $|\det S|= \frac{0}{0}$ [10], such that $|\det S(E_*\pm \epsilon)|=1.$  Here $\epsilon$ is arbitrarily small. Also the constancy of $|\det S|$ in (6) indicates its invariance under time-reversal as 
\begin{equation}
|\det S(-k)|=|\det S(k)|,
\end{equation} 
confirming the proposal (2). 

Nevertheless, let us point out that the possibility of CPA without lasing is due to the change of $|\det S(k)|$
under time-reversal for non-PT-symmetric potentials which in turn is due to (3). Thus, the following conditions may be met at  $k=k_c$
\begin{equation}
|\det S(k_c)|=0, \quad
T(-k_c)=R_{left}(-k_c) = 
R_{right}(-k_c)=\infty.
\end{equation}
and CPA alone (without lasing) is observed.
This is why coherent perfect absorbers are also called [9,15] time-reversed lasers. In this regard, one of the claim of this paper is that these potentials can not be PT-symmetric. 

We would like to remark that unlike the first proposal for the general CPA [9], the authors in [12] have been cautious about choosing the optical medium as P-symmetric. 
They set less general, yet simpler and intuitive condition for CPA at a real energy as $t+r_{left}=0=t+r_{right}$. For P-symmetric complex potentials the reciprocity ($r_{left}=r_{right}$) [1,3] works and that the result $\theta-\phi=\pi/2$ [22] of real Hermitian P-symmetric potentials is defied favorably due to the presence of non-Hermiticity (dissipation) so the CPA is feasible. This phenomenon has been called controlled CPA which is a special case of the more general condition (7) [9].

Thus, we conclude that complex PT-symmetric potentials do not display coherent perfect absorption alone (without lasing).
 
The above mentioned phenomena [7,9,10] occur in a complex potential as a possibility  and not as a necessity, therefore     
an exactly solvable is all the more welcome to bring out these
phenomena with their subtleties. The Scarf II  potential
\begin{equation}
V(x)=P \mbox{sech}^2 x + Q ~\mbox{sech} x \tanh x
\end{equation}
is a versatile potential entailing several interesting parametric regimes  in both PT-symmetric and non-PT-symmetric domains. 
By virtue of the  available beautiful 
complex transmission  and reflection amplitudes [2,23], Scarf II has helped in giving simple expressions for  SS
[8,14]. Recently, it has revealed [17] a rare (accidental) phenomena like reciprocity despite complex PT-symmetry and unitarity $(R+T=1)$ despite non-Hermiticity.
In the following we invoke three parametric regimes of complex Scarf II by complexifying  $P$ and $Q$ in various ways to demonstrate the novel [7,9,10] phenomena  coherent injection at optical potentials discussed above. 

We also study scattering from several  complex potentials by numerically integrating the Schr{\"o}dinger equation. However, here we discuss the results of two models. These models of complex potential are the rectangular
\begin{equation} 
V_R(x)=P \Theta_1(x)-iQ \Theta_2(x), \quad
\Theta_1(x)=\left\lbrace\begin{array}{lcr}
1, & &  |x|\le L\\
0, & &  |x|>L\\
\end{array}
\right.,\quad
\Theta_2(x)=\left\lbrace\begin{array}{lcr}
0, & & |x|\ge L\\
-1, & & -L < x <0\\
1, & & 0 \le x <L \\
\end{array}
\right.
\end{equation}
profile of compact support and the asymptotically converging Gaussian
\begin{equation}
V_G(x)=P e^{-x^2} + iQ x e^{-x^2},
\end{equation} 
with $P$ and $Q$ as parameters. In the following, we present the results of coherent scattering from  complex potentials (10-12) in three domains \{A,B,C\} in the light of the Scarf II potential.
\\ \\
\noindent
{\bf \{A\}:  Occurrence of CPA alone (non-PT-symmetric domain)}\\ \\
Let us consider the non-PT-symmetric domain of  Scarf II as
\begin{equation}
V_d(x)=(d^2-id)\mbox{sech}^2 x, \quad d\in R.
\end{equation}
This is an absorptive P-symmetric potential($d>0$) and it would be ideal for demonstrating  controlled CPA [12]. Using the transmission and reflection amplitudes [2,23] and by eliminating the Gamma functions (with complex argument) in them, in this  case (13), we find  
\begin{equation}
T(k)=\frac{k-d}{k+d}~ \frac {\sinh^2 \pi k} {\cosh^2 \pi k-\cosh^2 \pi d}.
\end{equation}
\begin{equation}
f_{left}(k)=-\frac{\sinh \pi d} {\sinh \pi k}=f_{right}(k),
\end{equation}
the reflection amplitudes are calculated as $r(k)=t(k)f(k)$,
the equivalence of left/ right in (15)  is by virtue of the P-symmetry of the potential (13) see [1,3]. Next we derive
\begin{equation}
|\det S(k)|=\left |\frac{k-d}{k+d} ~\frac{\cosh^2 \pi k-\cosh^2 \pi d}
{\cosh^2 \pi k-\cosh^2 \pi d}\right|.
\end{equation}
Notice that in (14), $k=-d$ is a pole (SS) and $k=+d$ is not a pole. So here $ k_c=d$ and the energy at which CPA occurs is $E_c=k_c^2=d^2$. We  use L'Hospital rule to see this
limit$_{k\rightarrow k_c} T(k)= \frac{\tanh \pi d}{4 \pi d}$,
which is finite. So there is only one SS.
The conjectured [14] properties (3) can be verified here readily.
More interestingly at $k=k_c$, $|\det S(k_c)|$ becomes indeterminate ($\frac{0}{0}$) 
but limit$_{k\rightarrow k_c}|\det S(k)|=0,$
Unlike the case of CPA with lasing [10] where it ought to be 1. This however presents the scenario of CPA [9]. The second aspect of CPA is met by noticing that $k=-k_c$ is clearly
an SS [equivalently SS at $k=k_c$ in time-reversed transmittance, $T(-k)$ (14)]. Figure 1 represents results (14,16) for Scarf II (13), which have also been recovered by numerical integration of
Schr{\"o}dinger equation, to confirm our numerical method.

We confirm the existence of CPA without lasing in non-PT-symmetric domains of two numerically solved complex potentials  (rectangular and Gaussian: see Figs 2,3). However, here in these models one has to carry out a judicious search of potential parameters to observe CPA. Like we find that for rectangular model for $P=2.21-1.091i, Q=0,L=2$ CPA exists at $E=E_c=4.015$ where $|\det S(E_c)|=0$ at this energy the spectral singularity occurs in the time-reversed transmission co-efficient, $T(-k_c)$. For the Gaussian model (12), we find that for $P=3.89 - 2.04i, Q=0$  the similar scenario, this time $E_c=3.992$.
CPA provides a crucial physical significance to the spectral
singularity wherein an SS in $T(k)$ at $k=\pm k_c$ implies CPA
at $k=\mp k_*$ at a real discrete energy $E=E_c$. Hence the coherent perfect absorbers are called time-reversed lasers [9,15].
 \\ \\
\noindent
{\bf \{B\}: The occurrence of CPA with lasing (broken PT-symmetry) }\\  \\
However, the situation changes dramatically when there is a spectral singularity present in the potential. Now let us consider the following parametrization of Scarf II potential
for $c \in R$
\begin{equation}
V_{c}(x)=[2c^2-1/4] \mbox{sech}^2 x - i[2c^2+1/2] \mbox{sech} x \tanh x. 
\end{equation}
For this case, we obtain
\begin{equation}
T(k)=\frac{\sinh^2 \pi k \cosh^2 \pi k} {(\cosh^2 \pi k- \cosh^2 \pi c)^2}. 
\end{equation}
and
\begin{small}
\begin{equation}
f_{left}(k)=i~[e^{-\pi k}-e^{\pi k} \cosh 2\pi c]~\mbox{cosech} 2\pi k, \quad  f_{right}(k)=i~[e^{\pi k}-e^{-\pi k} \cosh 2\pi c]~\mbox{cosech} 2\pi k.
\end{equation} 
\end{small}
One can readily notice self-dual SS [15] in transmission co-efficient (18) (poles at $k=\pm c$), i.e., at $E=c^2$ both $T(-c)$ and $T(c)$ are infinity. $r(k)$ is calculated as $r(k)=f(k)t(k)$. Next, using (18,19) in Eq. (5), we obtain
\begin{equation}
|\det S|=\frac{(\cosh^2 \pi k- \cosh^2 \pi c)^2} {(\cosh^2 \pi k- \cosh^2 \pi c)^2}. 
\end{equation}

Thus $|\det S|$ becomes indeterminate ($\frac{0}{0}$) at $k=\pm c$ meaning that $|\det S|=1$, for $k \ne \pm c$ and limit$_{k\rightarrow \pm c} |\det S| =1.$ This completes the explicit and the simplest  demonstration of the phenomenon called  CPA with lasing [10].

In Fig. 4, the scenario of CPA with lasing is shown for $c=2$, here we present the results due to numerical integration of the Scarf II potential. Notice a kinky behaviour  in $|\det S(E)|$ at
$E=E_*=c^2=4$ representing the indeterminacy. However, in  the neighbourhood of this energy, $|S|=1$ is retained.

In Fig. 5, we take rectangular (11) potential with $P=2.7, Q=-0.9, L=2$. Here the kinky behaviour in $|S(k)|$ (in (a)) and common spectral singularity in $T(K)$ and $T(-k)$ are displayed at $E_*=3.448$. For the Gaussian potential (12), for $P=4.0, Q=-6.25$, we get $E_*=3.380$  and the same scenario in Fig. 6, excepting that the kinky behaviour in Fig. 3(a) gets depicted as merely a dot at $E= 3.380$. The  indeterminacy of $|\det S(k)|$ at $E=E_*$ depicted as a kinky behaviour in Figs. 4(a), 5(a), 6(a) is the most subtle feature and is displayed the best by Scarf II potential  analytically (18-20) and not displayed so well by the numerical computation presented graphically (see Fig. 4(a)).
  \\ \\
\noindent
{\bf \{C\}: Non-occurrence of spectral-singularity and CPA (un-broken PT-symmetry)}\\ \\
When in (12), $P=-V_1, V_1>0$ and $Q=iV_2$ (both $V_1,V_2 \in R$ ), it has been shown [20] that if $|V_2| \le V_1+1/4$, the potential entails real discrete spectrum wherein the energy eigenstates are also eigenstates PT (PT-symmetry exact(unbroken)[20]), otherwise the real discrete eigenvalues disappear and make transition to non-real complex conjugate pairs and the PT symmetry is said to be spontaneously broken. Therefore, for all real values of  $a,b$ the potential
\begin{equation}
V_{a,b}(x)=-(a^2+b^2+a)\mbox{sech}^2 x-ib(2a+1) \mbox{sech} x \tanh x 
\end{equation}
can be verified to have finite number of real discrete eigenvalues and  the PT-symmetry remains unbroken.
Using the  available scattering amplitudes  [2,23], the following results follow from there [17].
\begin{equation}
T(k)=\frac{\sinh^2 \pi k \cosh^2 \pi k} {(\sinh^2 \pi k+ \sin^2 \pi a)(\sinh^2 \pi k+ \cos^2 \pi b)},
\end{equation}
and
\begin{equation}
f_{a,b}(k)= i \left [-\frac{\cos \pi a \sin \pi b}{\cosh \pi k} + \frac{\sin \pi a \cos \pi b}{\sinh \pi k} \right ].
\end{equation}
\begin{equation}
R_{left}(k)=T(k) |f_{a,b}(k)|^2, R_{left}(k)=T(k) |f_{a,-b}(k)|^2
\end{equation}
Verify that  the reflection and transmission (24) coefficients have common relevant poles at real discrete energies:
\begin{equation}
E_n=-(n-a)^2, E_m=-(m-1/2-b)^2,
\end{equation}
where $0\le n<a$ and $0 \le m < b+1/2$
which are two branches of the well known discrete eigenvalues [21,24]
of (21). The invariances given in (2) can be readily checked using (22-24) Further, we can write 
\begin{equation}
|\det S(k)|=T(k)~[1-f_{a,b}(k)f_{a,-b}(k)].
\end{equation}
Using (22,24), we eventually find that
\begin{equation}
|\det S(k)|= \frac{\sinh^4 \pi k+\sinh^2 \pi k(\sin^2 \pi a+ \cos^2 \pi b)+\sin^2 \pi a \cos^2 \pi b} {(\sinh^2 \pi k+ \sin^2 \pi a) (\sinh^2 \pi k+ \cos^2 \pi b)}=1.
\end{equation}
One can at once check that $T(k)$ (24) does not have any pole at a real $k$ and it can not become infinity (absence of SS) at any positive or negative real value of $k$. In several other numerically complex complex PT-symmetric potentials which possess real discrete spectrum we have found absence of SS and
hence CPA.
 
We would like to summarize our findings in the following:

$\bullet$ It is interesting to see that Hermitian and complex PT-symmetric potentials share yet another common feature that is $|\det S(E)|=1$ (6) for coherent scattering at them. However, the novel dissimilarity arises at $E=E_*$ (spectral singularity ) in the latter, where $|\det S|={0 \over 0}$ (indeterminate) and at this energy there occurs [10] CPA and lasing simultaneously giving rise to new types of lasers. 

$\bullet$ In  Figs 4,5,6 the parts (b) and (c) for various complex scattering potentials confirm our previous [14] conjecture that for complex PT-symmetric potentials or domains, $T(-k)=T(k)$ (this has been proved recently [19]).  

$\bullet$ Importantly, it turns out that apart from its analytic amenability as displayed amply in the new expressions (14-20, 27), the complex Scarf II potential is no way special. These (11,12) numerically solved  potentials behave qualitatively similarly in bringing out CPA (see Figs. 2-3) and CPA with lasing (see Figs. 5-6).

$\bullet$ As discussed in above here, a complex non-Hermitian potential can  have three parametric domains: \{A\}: non-PT-symmetric, \{B\}: PT-symmetric with broken PT-symmetry and \{C\}: PT-symmetric with unbroken PT-symmetry. Not undermining  the novel proposals and revelations of spectral singularity [7], CPA [9] and CPA with lasing [10], we would like to add that the necessary domain(s) of the potential could not be 
pinpointed due to the intuitive nature of these proposals. We elaborate this in the following. 

For example, the spectral singularity was proposed [7] for complex non-Hermitian potentials, here we find that  spectral singularity occurs only in \{A\} and \{B\} domains and does not occur in the \{C\} domain. CPA has been claimed [9] to occur in non-Hermitian potentials, here we have argued and found that CPA can not occur in complex  PT-symmetric domain owing to the result brought out here: $|S|=1$ (6). CPA with lasing was proposed [10] as a property of complex PT-symmetric potentials, later it was found [11] that it is actually the property of the broken PT-symmetric (\{B\} domain).

Lastly, we hope that the analytic amenability of Scarf II for studying coherent scattering at a complex potential has been well noted. We conclude that coherent perfect absorption can {\bf not} occur in complex PT-symmetric potentials. We have also conjectured that when PT-symmetry is unbroken (potential has real discrete spectrum) spectral singularity (and hence CPA with or without lasing) does not arise. This, however, requires a proof. We hope  that our work presented here strengthens the recent, novel and intuitive concepts of the wave propagation through non-Hermitian complex mediums/potentials.

\begin{figure}[H]
\centering
\includegraphics[width=5 cm,height=3.5 cm]{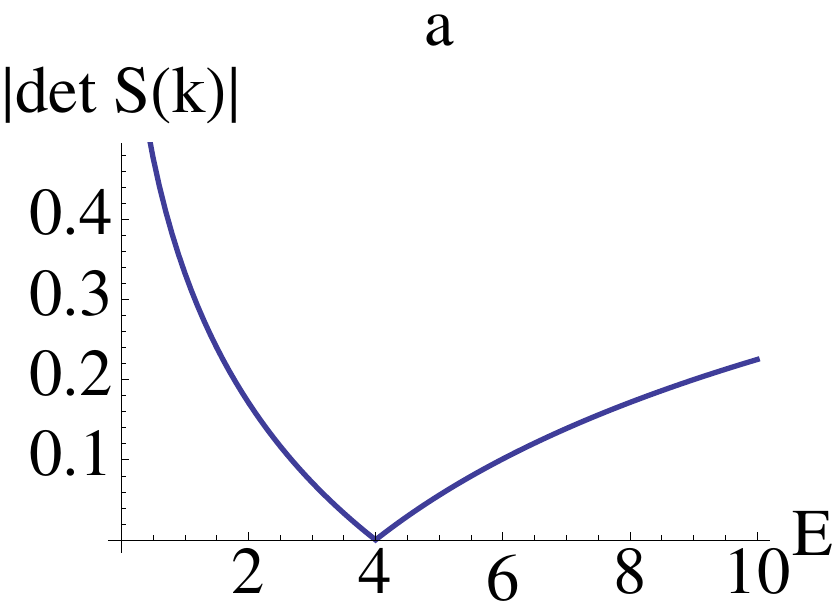}
\hskip .5 cm
\includegraphics[width=5 cm,height=3.5 cm]{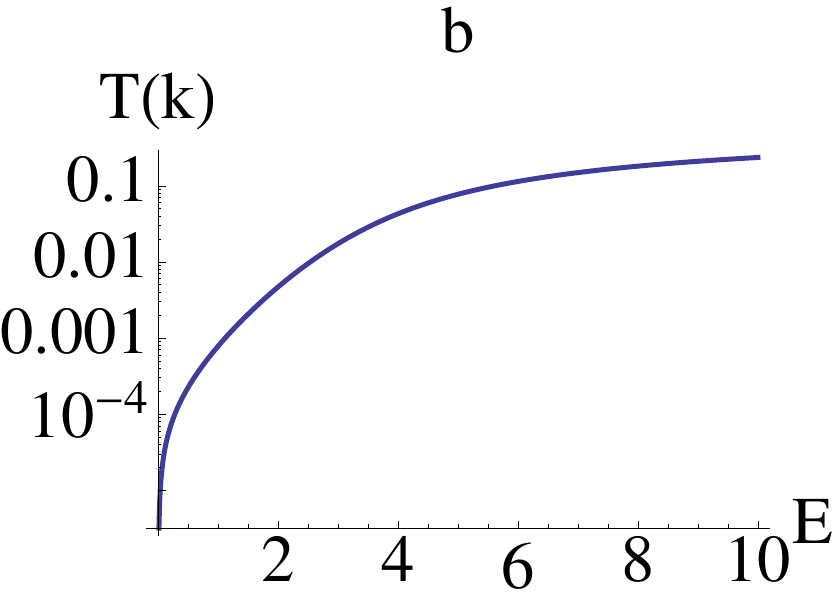}
\hskip .5 cm 
\includegraphics[width=5 cm,height=3.5 cm]{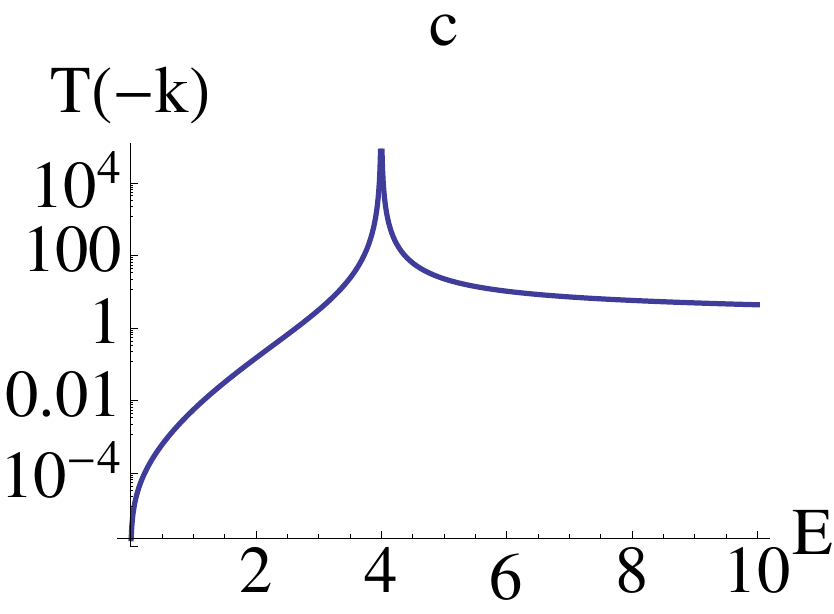}
\caption{The scenario of CPA without lasing for the complex (non-PT-symmetric) potential in Eq. (13) when the potential parameter $d=2$. In (a) the modulus of the determinant of the $2 \times 2$ s-matrix (5) of coherent scattering, (b) transmittance, (c) time-reversed transmittance
are plotted as a function energy.  Here, $E_c=4$, notice that $|S(E=E_c)|=0$ and the time-reversed transmittance ($T(-k)$ shows spectral singularity at $E=E_c$, whereas $T(k)$ is normal($<1$). Here $k=\sqrt{E}.$ }
\end{figure}
\begin{figure}[H]
\centering
\includegraphics[width=5 cm,height=3.5 cm]{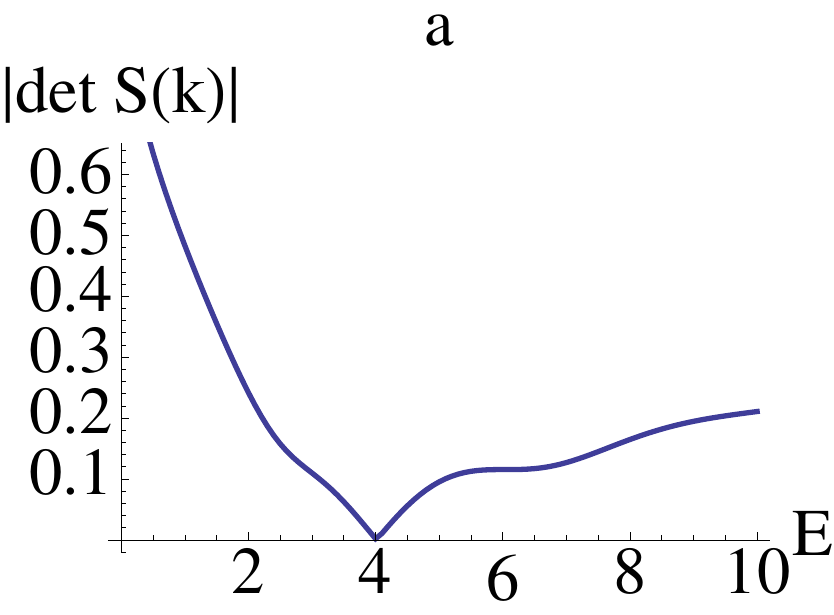}
\hskip .5 cm
\includegraphics[width=5 cm,height=3.5 cm]{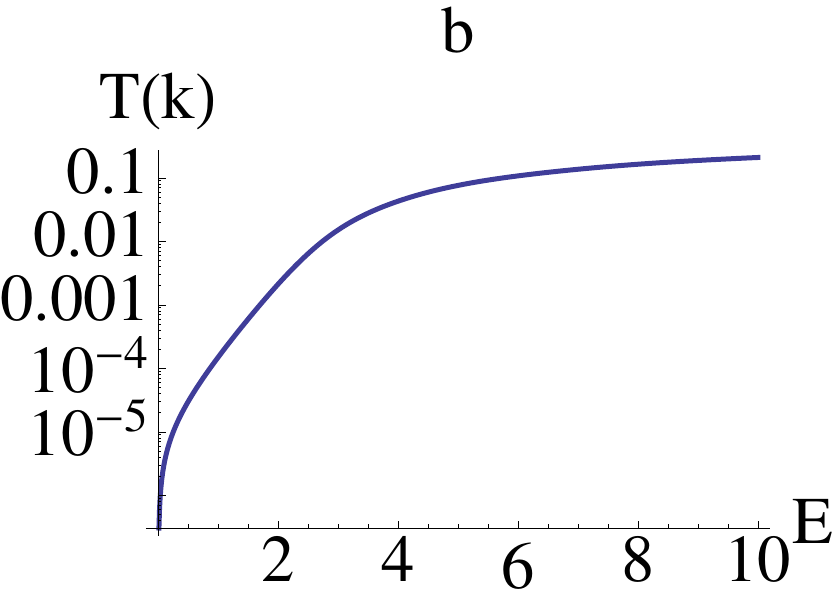}
\hskip .5 cm 
\includegraphics[width=5 cm,height=3.5 cm]{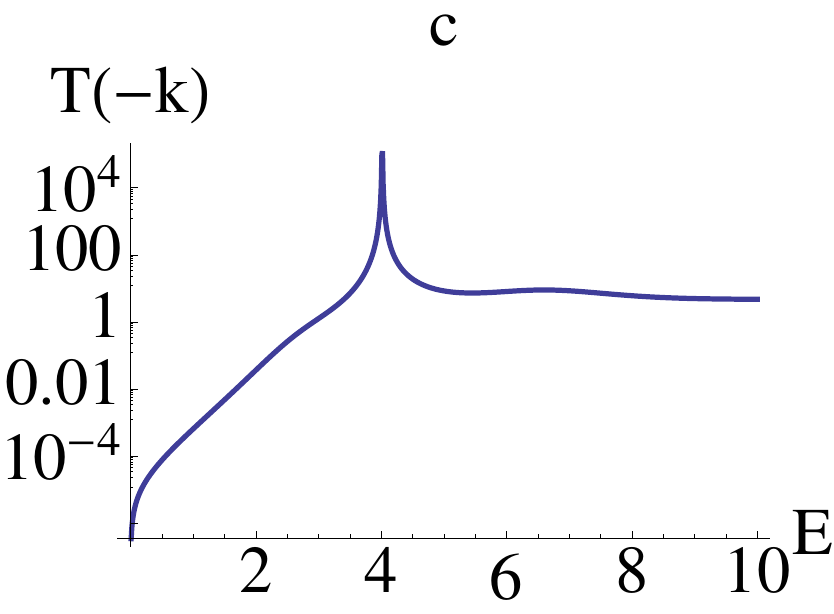}
\caption{ The same as in Fig. 1. Here the complex (non-PT-symmetric) potential is rectangular (Eq. (11): when $L=2, P= 2.21-1.09i, Q=0$) and $E_c=4.015$.}
\end{figure}
\begin{figure}[H]
\centering
\includegraphics[width=5 cm,height=3.5 cm]{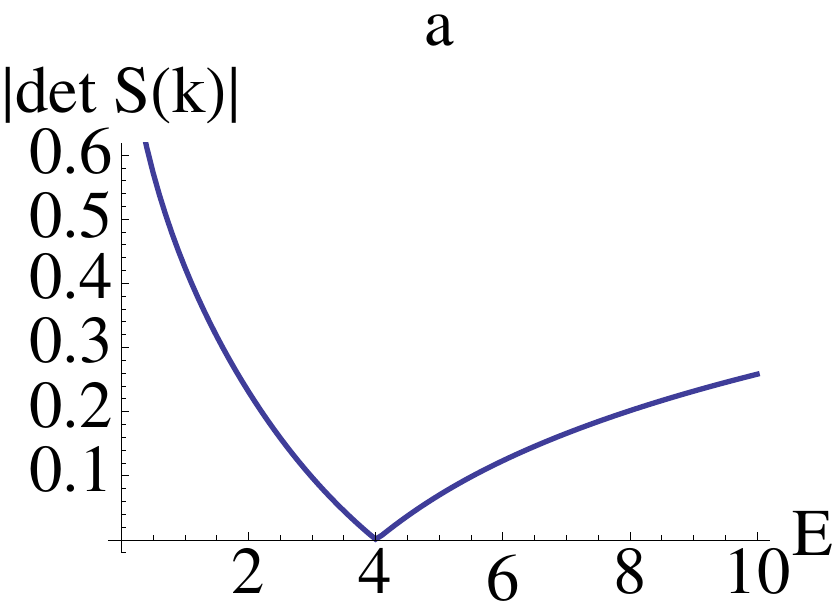}
\hskip .5 cm
\includegraphics[width=5 cm,height=3.5 cm]{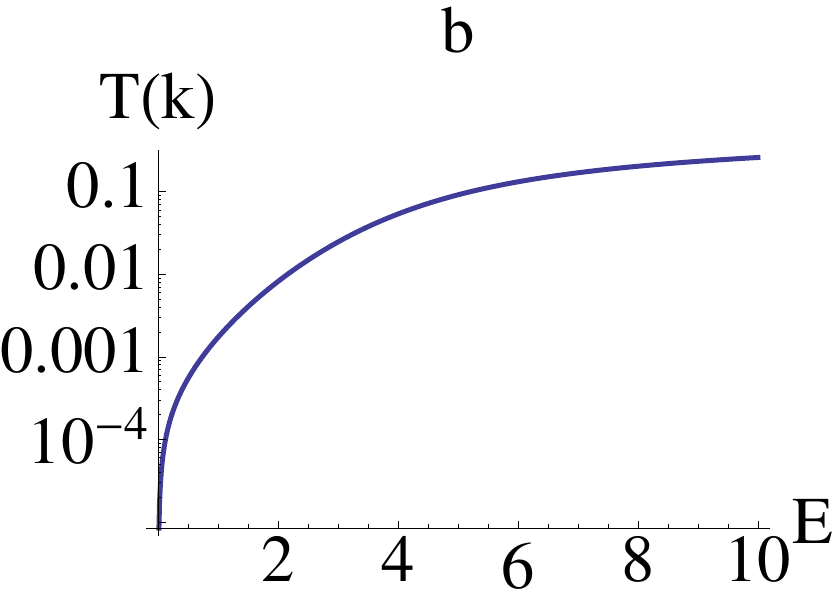}
\hskip .5 cm 
\includegraphics[width=5 cm,height=3.5 cm]{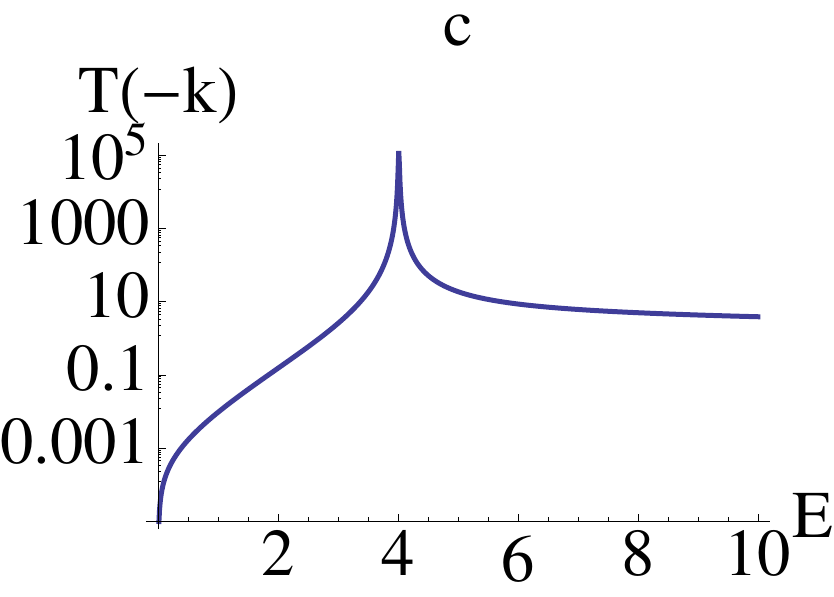}
\caption{The same as in Fig. 1,2 excepting that here the potential is Gaussian (see Eq. 12: $P= 3.89-2.04i, Q=0$) and $E_c=3.992$ }
\end{figure}
\begin{figure}[H]
\centering
\includegraphics[width=5 cm,height=3.5 cm]{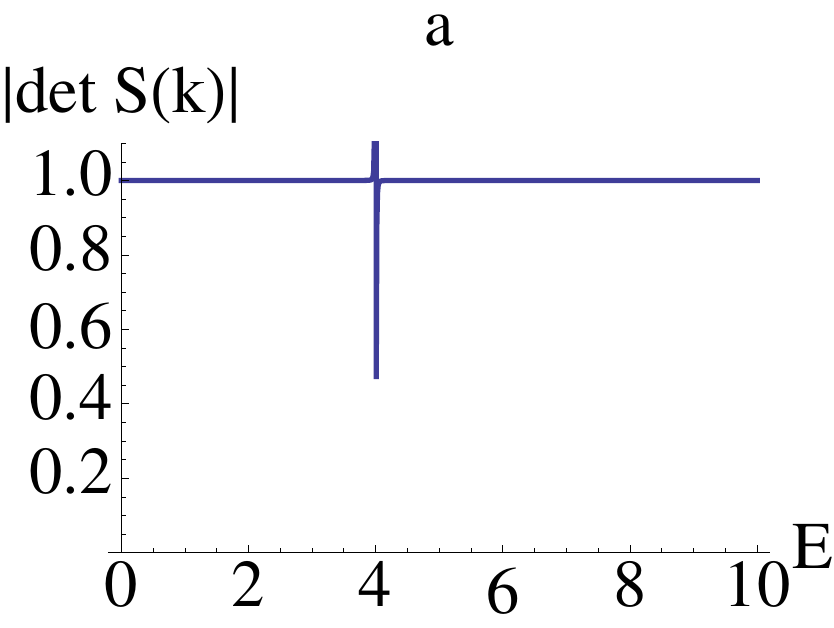}
\hskip .5 cm
\includegraphics[width=5 cm,height=3.5 cm]{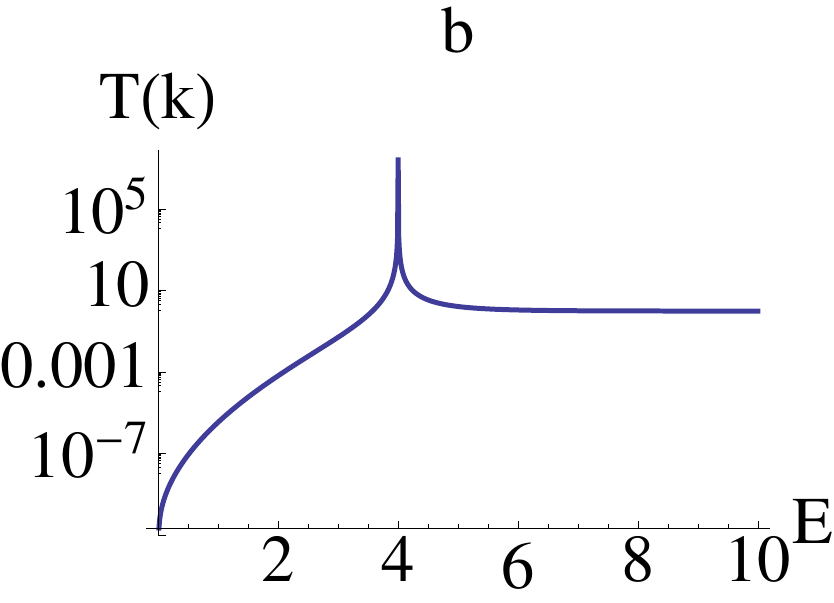}
\hskip .5 cm 
\includegraphics[width=5 cm,height=3.5 cm]{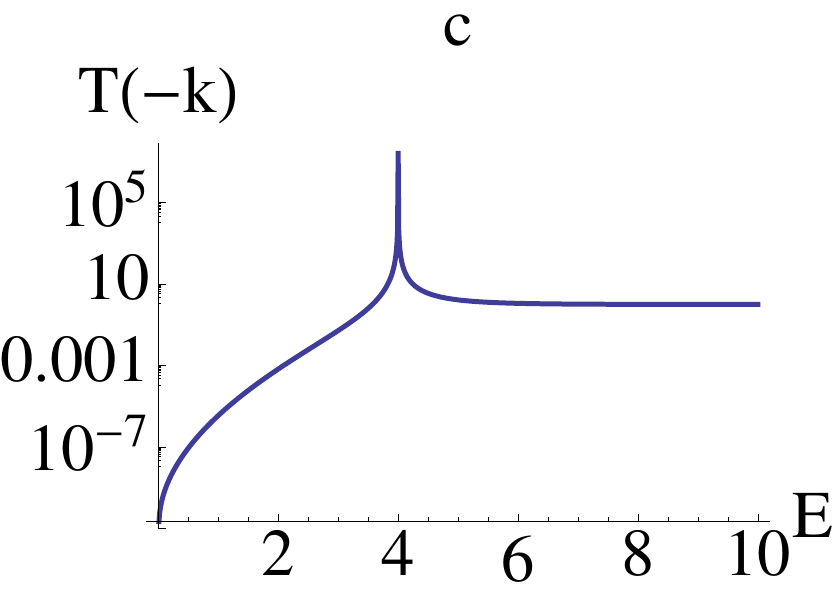}
\caption{The scenario of CPA with lasing for the complex PT-symmetric potential in Eq. (17) when $c=2$. This is the domain of broken PT-symmetry ($V(x)$ does not possess real discrete spectrum). Both $T(k)$ and $T(-k)$ are identical conforming to Eq.(2) as conjectured in [14]. Here $E_*=4$, notice that in (a) $|\det S(E_*)|$ shows indeterminacy by admitting some spurious value other than 1 appearing as a spike. However at energy very close to $E=E_*$ and at other energies the value 1 is attained.
This is the characteristic of simultaneous occurrence of  CPA and lasing in a potential.  This, however, is the best displayed by Eq.(20), so far.}
\end{figure}
\begin{figure}[H]
\centering
\includegraphics[width=5 cm,height=3.5 cm]{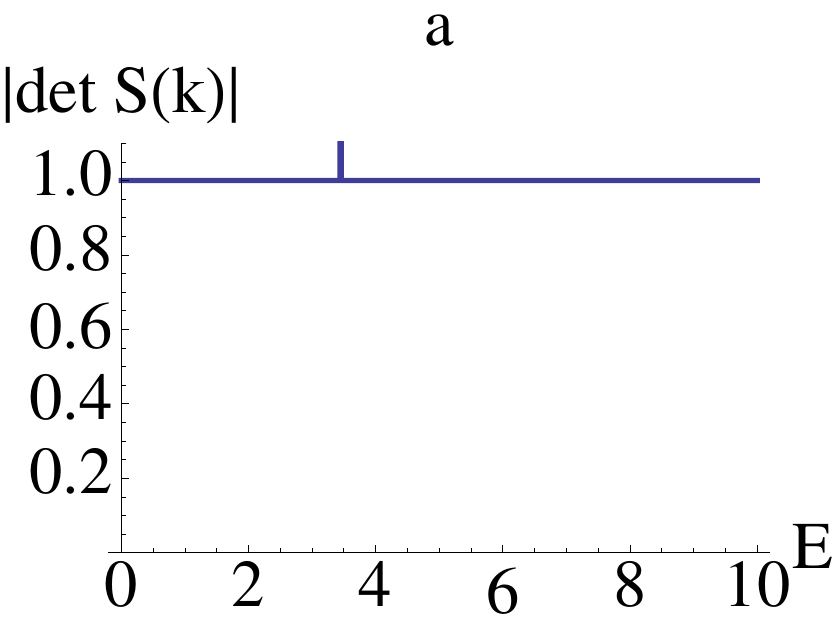}
\hskip .5 cm
\includegraphics[width=5 cm,height=3.5 cm]{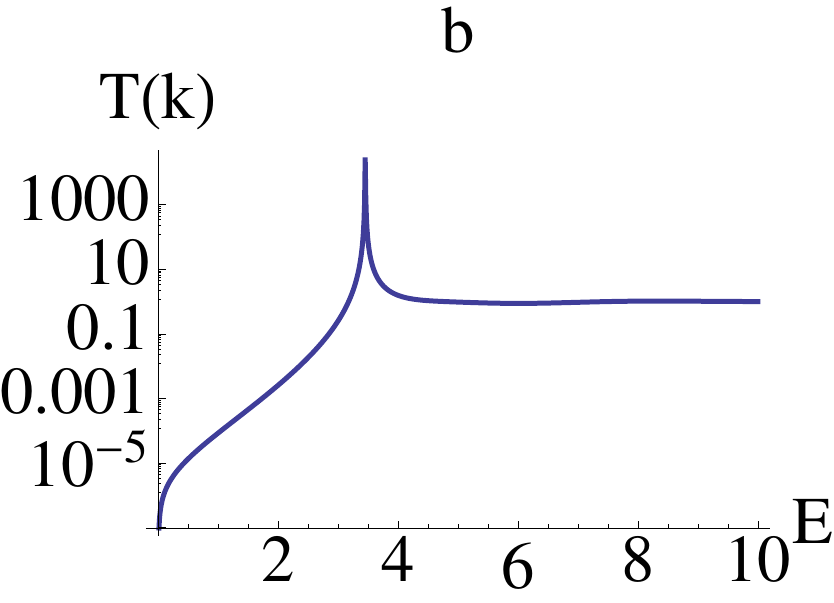}
\hskip .5 cm 
\includegraphics[width=5 cm,height=3.5 cm]{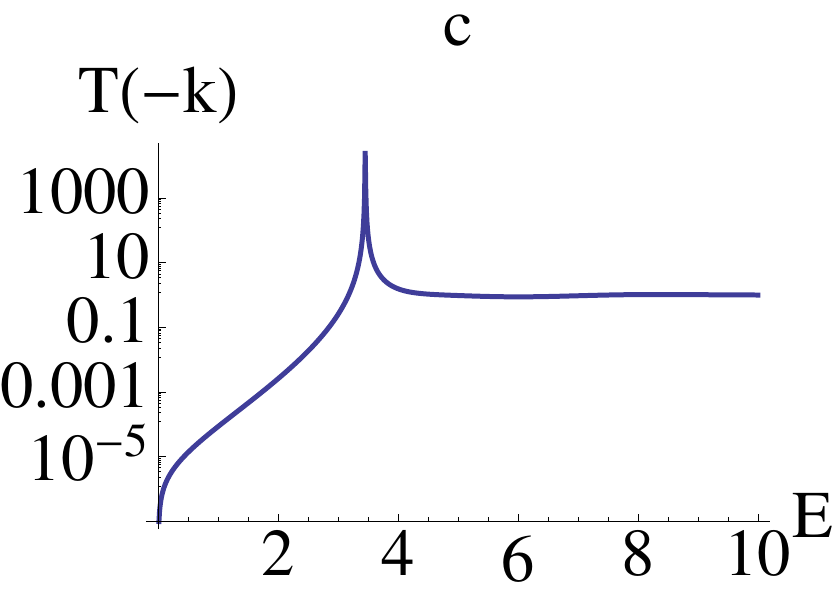}
\caption{The same as in Fig. 4, Here we have complex
PT-symmetric rectangular model : $ P=2.7, Q=-0.9, L=2 $ in Eq.(11). Here $E_*=3.448$, notice that in (a) $|\det S(E_*)|$ is a spurious value other than 1 appearing as a kink. The spectral singularity occurs at the energy $E=E_*$ in both $T(k)$ and $T(-k)$.}
\end{figure}
\begin{figure}[H]
\centering
\includegraphics[width=5 cm,height=3.5 cm]{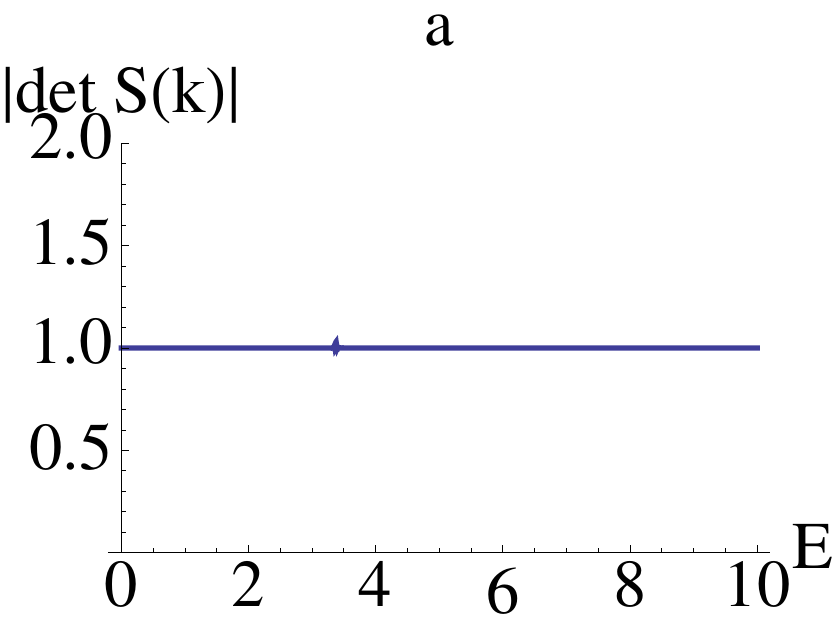}
\hskip .5 cm
\includegraphics[width=5 cm,height=3.5 cm]{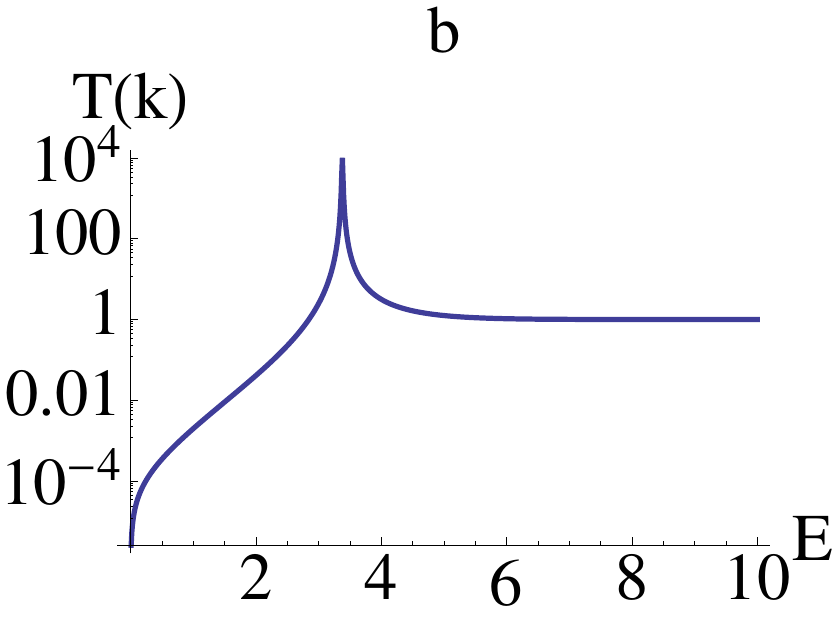}
\hskip .5 cm 
\includegraphics[width=5 cm,height=3.5 cm]{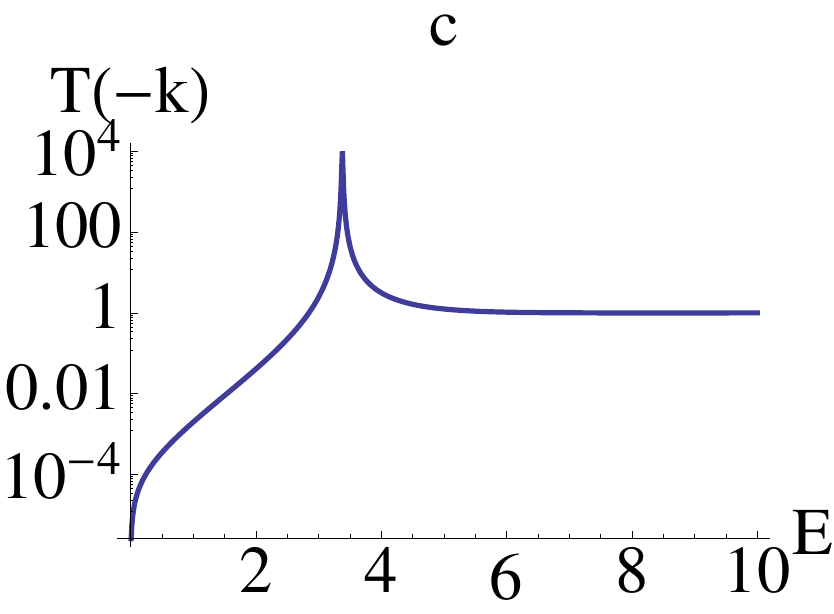}
\caption{The same as in Fig. (4,5) for complex PT-symmetric Gaussian potential: (Eq.(12) with $P= 4.0, Q=-6.25$). Here, $E=3.380$ and notice in (a) $|\det S(E_*)|$ is a spurious number other than 1 appearing as a small vertical line}
\end{figure}

\end{document}